\documentclass{article}
\usepackage{spconf,amsmath,graphicx, amsfonts, hyperref}
\usepackage{mathtools,tikz,caption}
\usepackage{booktabs}
\DeclareRobustCommand\sampleline[1]{%
  \tikz\draw[#1] (0,0) (0,\the\dimexpr\fontdimen22\textfont4\relax)
  -- (1.2em,\the\dimexpr\fontdimen22\textfont4\relax);%
}
\def\x{{\mathbf x}}
\def\L{{\cal L}}

\title{How Redundant Is the Transformer Stack in \\Speech Representation Models?}

\name{Teresa Dorszewski$^*$, Albert Kj\o ller Jacobsen$^*$, Lenka T\v{e}tkov\'a, Lars Kai Hansen \thanks{$^*$ Equal contribution. \\This
work was supported by the Pioneer Centre for AI, DNRF grant number P1, the DIREC Bridge project Deep Learning and Automation of Imaging-Based Quality of Seeds and Grains, Innovation Fund Denmark grant number 9142-00001B, and the Novo Nordisk
Foundation grant NNF22OC0076907 ”Cognitive spaces - Next generation explainability”. \\
\copyright 2025 IEEE. Personal use of this material is permitted. Permission from IEEE must be obtained for all other uses, in any current or future media, including reprinting/republishing this material for advertising or promotional purposes, creating new collective works, for resale or redistribution to servers or lists, or reuse of any copyrighted component of this work in other works.}}
\address{Technical University of Denmark\\ DTU Compute, Section for Cognitive Systems \\ \texttt{\{tksc,akjja,lenhy,lkai\}@dtu.dk}}

\begin{document}
\ninept
\maketitle
\begin{abstract}
Self-supervised speech representation models, particularly those leveraging transformer architectures, have demonstrated remarkable performance across various tasks such as speech recognition, speaker identification, and emotion detection. Recent studies on transformer models revealed high redundancy between layers and the potential for significant pruning, which we will investigate here for transformer-based speech representation models.
We perform a detailed analysis of layer similarity in speech representation models using three similarity metrics: cosine similarity, centered kernel alignment, and mutual nearest-neighbor alignment. Our findings reveal a block-like structure of high similarity, suggesting two main processing steps and significant redundancy of layers. We demonstrate the effectiveness of pruning transformer-based speech representation models without the need for post-training, achieving up to 40\% reduction in transformer layers while maintaining over 95\% of the model’s predictive capacity. Furthermore, we employ a knowledge distillation method to substitute the entire transformer stack with \textit{mimicking layers}, reducing the network size by 95-98\% and the inference time by up to 94\%. This substantial decrease in computational load occurs without considerable performance loss, suggesting that the transformer stack is almost completely redundant for downstream applications of speech representation models.  
\end{abstract}

\section{Introduction}
\label{sec:intro}
Recent transformer-based speech representation models have shown impressive performance in numerous tasks including speech recognition, speaker identification, and emotion detection \cite{superb2021, chen2022wavlm, baevski2020wav2vec}. However, these models often come with significant computational costs due to their large size and complexity. This paper investigates the redundancy present within transformer layers of speech representation models, exploring the potential to prune or replace these layers and thereby create smaller, more efficient networks suitable for on-device automatic speech recognition tasks. 

Several studies have shown that transformer-based speech representation models contain a substantial amount of redundancy~\cite{liu2021tera, peng2023structured, zhang2021usefulness, sajjad2023layers}. Recent research on large language models (LLMs) has revealed that many layers and neurons can be pruned without significantly impacting performance~\cite{dalvi2020analyzing, yang2024laco, gromov2024unreasonable, men2024shortgpt}. Similar findings have been observed in speech representation models, where pruning or informed layer selection can lead to reduced computational requirements and faster inference times while retaining or even improving performance \cite{pasad2023comparative, dorszeswki2024pruning}.

Moreover, a high degree of linearity was observed in transformer models, further indicating potential redundancy~\cite{razzhigaev2024your}. It was demonstrated that the embedding transformations between sequential layers exhibit near-perfect linearity, suggesting that many of these layers may be performing redundant operations. By identifying and removing the most linear layers or replacing them with linear approximations, they show that it is possible to remove a few layers without loss in performance.

Recent studies on knowledge distillation (KD) of speech representation models have shown the potential to significantly reduce the number of parameters while mostly maintaining performance in many downstream tasks \cite{yang2022knowledge, peng2023dphubert, huang2023ensemble, zampierin2024skill}. Zampierin et al. \cite{zampierin2024skill} utilize a similarity-aware strategy, leveraging redundancy of layers to reduce the number of layers needed. 

With this paper, we aim to investigate this redundancy systematically and leverage it in a pruning and KD approach for speech representation models. Our main contributions include:

\textbf{(1)} A detailed analysis of similarity in speech representation models, leveraging three similarity metrics. We find high similarity between layers that presents itself in a block-like structure, suggesting two main processing steps and high redundancy of layers. 

\textbf{(2)} Evidence of the effectiveness of pruning transformer-based speech representation models without the need for post-training. We can remove up to 45\% of the transformer layers without significant loss in performance by structurally selecting layers leveraging the block influence score \cite{men2024shortgpt}. We find that to maintain performance, parts of both blocks identified using similarity metrics need to be present. 

\textbf{(3)} Significant reduction in the computational footprint of transformer-based speech representation models while maintaining 95\% of the models' predictive capacity. By employing a knowledge distillation approach and replacing the whole transformer stack with \textit{mimicking layers}, we can decrease the network size by an order of magnitudes. 

Our experiments consistently show high redundancy in the transformer layers of speech representation models, indicating that the transformer stack can be considerably minimized for downstream applications to create more resource-efficient networks.

\begin{figure*}[t]
    \centering
    \includegraphics[width=\linewidth]{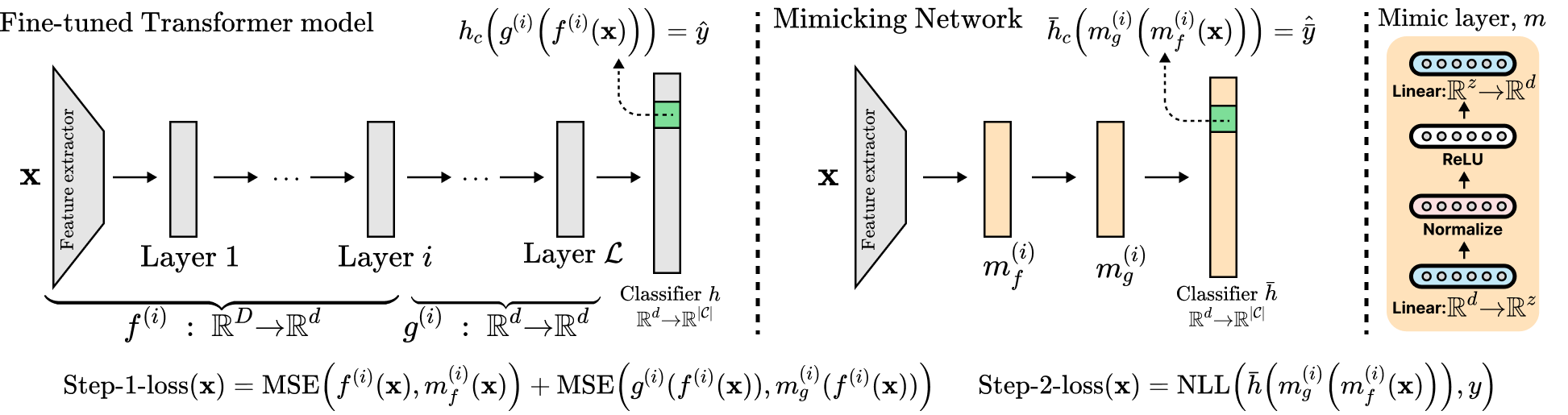}
    \caption{\textbf{Conceptual overview of mimicking networks.} \textit{\textbf{Left}:} The general transformer model, consisting of a feature extractor module, the transformer stack and a classification/probing layer. We denote the first part of the network until layer $i$ by $f^{\left( i\right)}$ the remaining part of the network from layer $i$ to $\mathcal{L}$ by $g^{\left( i \right)}$, and the classification layer by $h$. \textit{\textbf{Middle}:} A 2-layer mimicking network where $m_f^{\left(i\right)}$ and $m_g^{\left(i\right)}$ represent mimicked representations of $f^{\left(i\right)}$ and $g^{\left(i\right)}$, respectively, learned via the step-1-loss. In step 2, the classifier, $\overline{h}$, and mimicking layers are then finetuned for the downstream task. \textit{\textbf{Right}:} Design of the mimicking layer that maps an input embedding of dimension $d$ to a $z$-dimensional representation before mapping it back to the original shape.}
    \label{fig:model-overview}
\end{figure*}

\section{Methods}
\label{sec:methods}

We divide the transformer-based network into three parts. Let ${f^{(i)}:\mathbb{R}^D \rightarrow \mathbb{R}^d}$ for ${i\in\{1, \dots, \mathcal{L}\}}$ be the network up to layer $i$ (with embedding dimension $d$), let ${g^{(i)}:\mathbb{R}^d \rightarrow \mathbb{R}^d}$ be the remaining part of the network from layer $i$ to $\mathcal{L}$, and ${h:\mathbb{R}^d \rightarrow \mathbb{R}^{|\mathcal{C}|}}$ the classification layer. The composite function ${\left(h \circ \left( g^{(i)} \circ f^{(i)}\right)\right)\left(\boldsymbol{x}\right)}$ is then the complete forward pass 
for an input $\boldsymbol{x} \in \mathbb{R}^D$.
For the rest of this section, for simplicity we fix the notation: ${i, j\in\{1, \dots, \mathcal{L}\}}$ indices of transformer layers, ${X\in\mathbb{R}^{n\times D}}$ a matrix of input data, and ${A^{(i)} = f^{(i)}(X)\in \mathbb{R}^{n\times d}}$ representation of input after layer $i$.

\subsection{Layer Similarity}
First, we perform an extensive analysis of latent representations after each transformer layer of several speech representation models (see \autoref{sec:models}), where we investigate the similarity between layers to identify redundant information. We extract latent representations of audio input after each transformer block and compare the representations across layers, leveraging three similarity metrics, namely cosine similarity, centered kernel alignment (CKA), and mutual nearest neighbor alignment (mutual kNN). All metrics are invariant to isotropic scaling and orthogonal transformations implying permutation invariance. All scores depend on the input data $X$, which we omit in the notation for simplicity. Moreover, we center all the representations.

The cosine similarity score between representations of $X$ at layers $i$ and $j$ is defined as 
\begin{equation}
    \operatorname{S}_{cos}(i, j) = \frac{1}{n}\sum_{l=1}^n\frac{\left(A^{(j)}_{l, \cdot}\right)^T A^{(i)}_{l, \cdot}}{\left\Vert A^{(i)}_{l, \cdot} \right\Vert \left\Vert A^{(j)}_{l, \cdot}\right\Vert}.
\end{equation}

We additionally consider the CKA metric \cite{kornblith2019similarity}. In the linear form, CKA is given by 
\begin{equation}
\operatorname{S}_{CKA}(i, j) = \frac{\left\Vert\left( A^{(j)}\right)^T A^{(i)}\right\Vert_F^2}{\left\Vert\left( A^{(i)}\right)^T A^{(i)}\right\Vert_F \left\Vert\left( A^{(j)}\right)^T A^{(j)} \right\Vert_F }.
\end{equation}

Acknowledging the ongoing discussion on the validity of similarity metrics for learned representations \cite{huh2024platonic}, we consider the locally-aware mutual kNN similarity score given by
\begin{equation}
    \operatorname{S}_{kNN}\left(i, j\right) = \frac{1}{n} \sum_{l=1}^n \left(\frac{1}{k} \left|\mathcal{N}_k\left(A^{(i)}_{l, \cdot}\right) \cap \mathcal{N}_k\left(A^{(j)}_{l, \cdot}\right)\right|\right),
\end{equation}
where $\mathcal{N}_k\left(A^{(i)}_{l, \cdot}\right)$ is the set of indices for the $k$-nearest samples of $A^{(i)}_{l, \cdot}$ in the batch. We chose $k=8$ based on the robustness of initial experiments and previous studies \cite{huh2024platonic}. 

\subsection{Pruning}

We investigate the relation between feature similarity patterns and model redundancy by heuristically pruning the transformer stack. The considered heuristics include \textit{forward} and \textit{backward} pruning, i.e. pruning starting with the first or last transformer block, along with pruning by the minimum Block Influence (BI) score $\operatorname{BI}\left(i\right) = 1 - \operatorname{S}_{cos}(i-1, i)$ \cite{men2024shortgpt} and a modified, locally-aware version relying on mutual kNN similarity, $\operatorname{S}_{kNN}$, rather than cosine similarity, $\operatorname{S}_{cos}$. For all heuristics, we prune by deleting whole transformer blocks in the order determined by the heuristic, the first transformer block is never pruned. The remaining blocks are stitched together without any post-training.

\begin{figure*}[t]
    \centering
    \includegraphics[width=\linewidth]{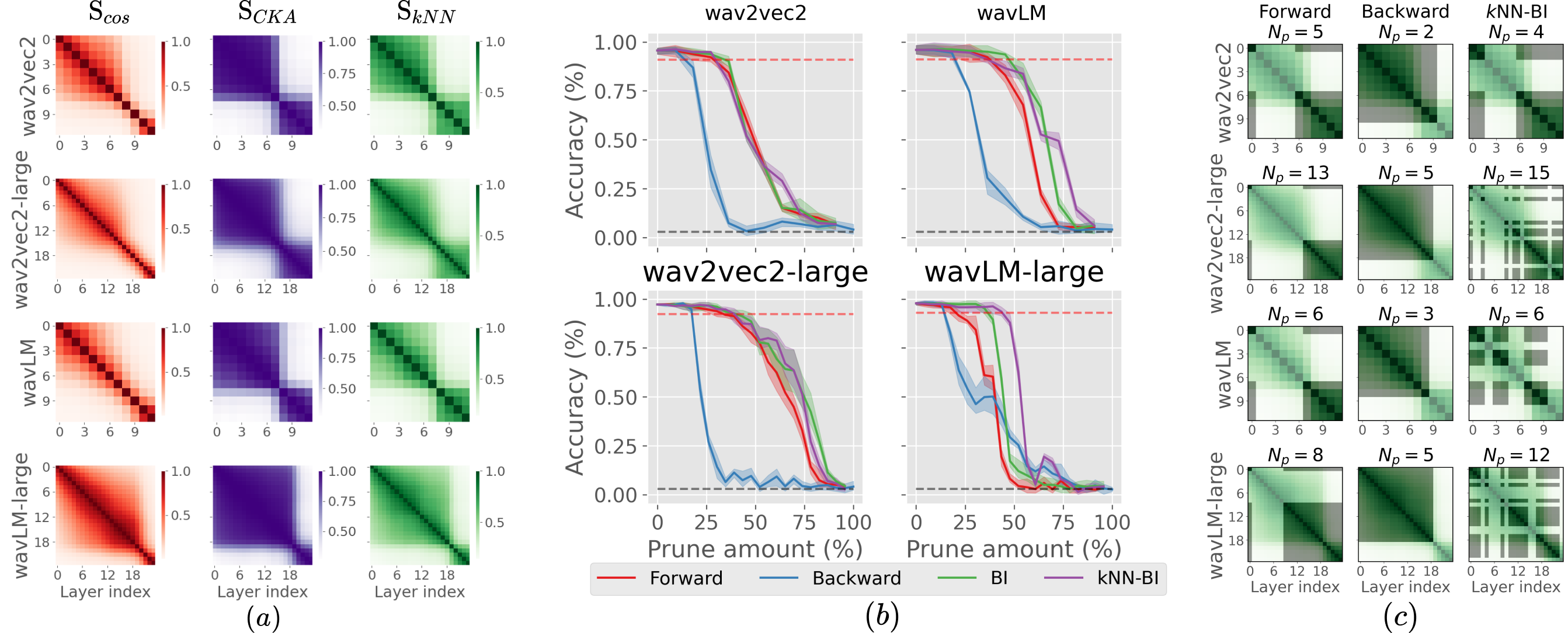}
    \caption{\textbf{Analysis of redundancy of layers} using similarity measures and pruning of layers: $(a)$ Similarity between layers of \texttt{wav2vec2} and \texttt{wavLM}. All three metrics (cosine similarity, CKA, mutual kNN) reveal a block structure. $(b)$ Effect of pruning on performance, using four different pruning objectives. Up to 45\% of layers can be pruned while maintaining 95\% of accuracy ( \sampleline{dashed, red}). After pruning most layers, the model performance drops to random chance ( \sampleline{dashed, gray}). Uncertainty ranges cover the empirical 2.5 and 97.5 quantiles obtained from $N=5$ runs. $(c)$ Visualisation of pruned layers on top of the kNN similarity matrix (with $50\%$ performance threshold). Light layers indicate pruned layers, dark layers are still present. Backward and forward pruning (left+middle) only preserve performance as long as both blocks are still present. Pruning based on kNN-BI (right) prunes layers mainly in the first block.}
    \label{fig:sim_and_pruning}
\end{figure*}

\subsection{Mimicking Networks - Knowledge Distillation}
We propose a simple strategy for distilling knowledge from fine-tuned audio models (teacher networks) based on reproducing intermediate representations. We introduce a so-called \textit{mimicking network}, which is trained to replace the transformer stack using 1 or 2 \textit{mimicking layers}. As illustrated in \autoref{fig:model-overview}, the mimicking network learns to reproduce representations in the last layer $\L$ of the teacher transformer stack and optionally also in the layer $i$. As experimental parameters, we consider the layer type, i.e. \textit{Transformer} (\texttt{TransformerEncoderLayer} from PyTorch) or \textit{linear mimic} layers (\autoref{fig:model-overview} right), as well as their hidden dimensionality, $z \in \{32, 768, 4096\}$. We ensure weight-sharing along the temporal dimension. Each model is trained with a 2-stage procedure consisting of \textit{1) a mimicking phase} using a mean squared error ($\operatorname{MSE}$) objective and \textit{2) an adaptation phase} for fine-tuning to the downstream task using a negative log likelihood ($\operatorname{NLL}$) objective on the log-probabilities (detailed loss function in \autoref{fig:model-overview}). For evaluating the mimicking approach, we additionally explore randomly initialized models that only consider the adaptation phase. 

In the \textit{mimicking phase}, models are trained for 50 epochs using the training set with a batch size of 128, resulting in 33150 steps. Model evaluation is carried out regularly on 1024 inputs randomly sampled from the validation set. All models converged within the training horizon. Subsequently, the \textit{adaption phase} exploits the optimal weight set and trains for 30 epochs, i.e. 19890 steps. The optimal models based on the validation set, i.e. before potential overfitting, are saved. Based on initial experiments, all models are trained with a fixed learning rate of $10^{-3}$. 

All models are evaluated on the test set by individually predicting the 4482 samples, and the inference time is measured after a GPU warm-up of 300 steps (to ensure comparable conditions). We report average performances and inference times including uncertainty estimates given by the standard error of the mean.

\subsection{Data \& Models}
\label{sec:models}
All analyses consider a word classification task using the \textit{speech commands v0.02} dataset \cite{speechcommandsv2}, which features 35 words (i.e. classes) spoken by more than 400 speakers. The dataset and its splits are obtained from \href{https://huggingface.co/datasets/google/speech_commands}{\textit{huggingface.co}}. All audio files are resampled to 16kHz and padded/restricted to 1 second, and the \textit{\_silence\_} class is excluded for the analyses.

We consider the \texttt{base} and \texttt{large} fine-tuned versions of \texttt{wav2vec2}~\cite{baevski2020wav2vec} and \texttt{wavLM}~\cite{chen2022wavlm}. All models follow the same transformer architecture \cite{vaswani2017attention} with 12 or 24 transformer stacks and were fine-tuned on the train set to perform classification of the 35 words, with a learning rate of $2\x10^{-5}$ for 10000 steps. The final accuracy on the test set of the models is 98.32 / 97.21\% for \texttt{wav2vec2} (base/large) and 97.22 / 98.86\% for \texttt{wavLM} (base/large).

\section{Results \& Discussion}
\label{sec:results}

\subsection{Layer Similarity}
First, we explore the similarity structure between layers. Our analysis reveals that all models exhibit two primary blocks characterized by highly similar latent representations throughout each block (see \autoref{fig:sim_and_pruning}a). The first block consists of approximately two-thirds to three-quarters of the transformer layers, while the second block typically comprises the final four to five layers. These findings suggest a significant degree of redundancy within these blocks, raising questions about the necessity of all layers.
Similar block structures have been observed (but not further investigated) in convolutional models \cite{kornblith2019similarity} and other speech representation models \cite{zampierin2024skill}, with varying numbers and dimensions of the blocks depending on the models and tasks.

When comparing the three similarity metrics, CKA and mutual kNN exhibit the block structure more distinctly than cosine similarity. Consistent with recent debates on similarity metrics \cite{kornblith2019similarity,huh2024platonic}, our findings indicate that CKA and mutual kNN more accurately capture similarity structures, and mutual kNN reveals additional details within the blocks. This further indicates the potential benefit of pruning according to local rather than global similarity structure.

\subsection{Layer-Wise Pruning}
Given the high similarity and therefore potential redundancy of the transformer layers, we investigate how many layers can be pruned, i.e. simply deleted without retraining, before significantly impacting performance. For all models, we can prune a substantial number of layers before we see a significant drop in performance (see \autoref{fig:sim_and_pruning}b). When using BI or kNN-BI to prune the least important layers first, we can prune 25-42\% of layers while maintaining 95\% of the original performance. Interestingly, when pruning forward (not pruning the very first layer), we see very similar results, where we can prune almost up to the same amount of layers, which indicates that especially early layers within the first block are redundant. 

When pruning backward or forward, the performance drops completely after removing either of the blocks identified earlier (with exception of wavLM large, where the performance already drops after removing half of the first block). In \autoref{fig:sim_and_pruning}c it becomes apparent that the performance is maintained only as long as parts of both blocks are still present, highlighting the importance of both processing steps.

Other studies have shown that by retraining after layer-wise pruning, models can regain full performance or, in some cases, even improve performance in speech representations models \cite{dorszeswki2024pruning} and LLMs \cite{fan2021layer}. This highlights how redundant layers are not only a computational burden, but are also unnecessary or even harmful to keep. 

\subsection{Mimicking Networks}

\begin{figure}
    \centering
    \includegraphics[width=\linewidth]{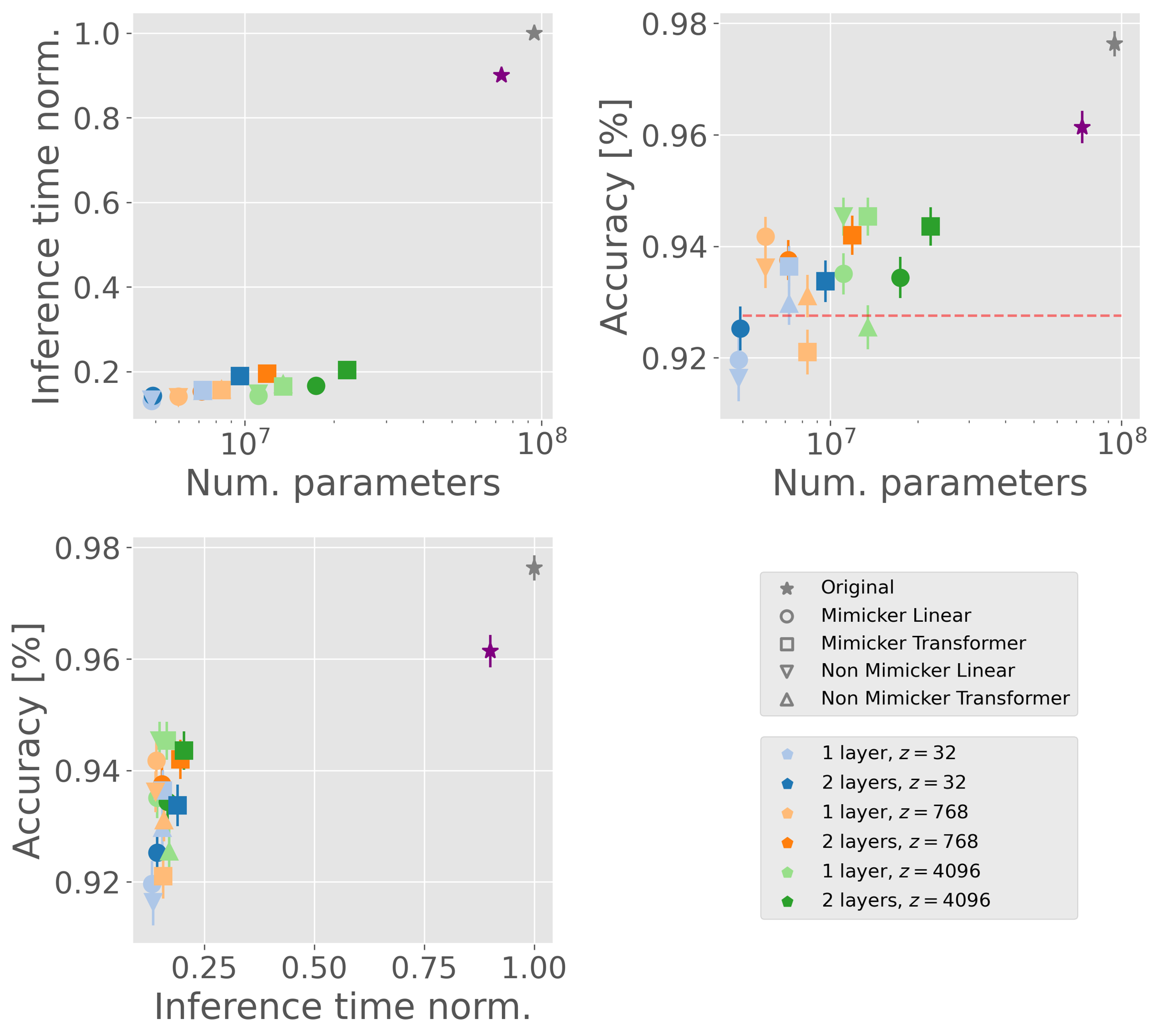}
    \caption{\textbf{Simplification of transformer stack using mimicking networks.} Reduction in inference time (up to 87\%) and number of parameters (up to 95\%) using mimicking networks, while maintaining 95\% of the original accuracy (\sampleline{dashed, red}). We test transformer and linear layers with different dimensions $z$. Inference time is normalized to 1 (inference time of original model), the pruned model has 3 layers pruned using kNN-BI. These results are for \texttt{wav2vec2}, results for other models are in the appendix.}
    \label{fig:mimic}
\end{figure}

Is the transformer stack completely redundant? To answer this question, we substitute the transformer stack with mimicking layers. Based on the two blocks we find during the similarity analysis, we first test two mimicking layers, with the intermediate mimicked layer $i$ being the last layer of the first block. We compare these results to only using one mimicking layer, directly learning the final representations, as well as to using just a linear or transformer layer instead, directly trained for classification. These substitutions lead to a parameter reduction of up to 95\% in base models and up to 98\% in large models and a reduction of the inference time of up to 94\% (see\autoref{fig:mimic} and \autoref{fig:mimic_wav2veclarge}-\ref{fig:mimic_wavlm-large} in the appendix). Most models retain more than 95\% of the performance while achieving these high reductions in size and therefore also inference time. This is a higher reduction, while maintaining comparable performance, than in recent studies on KD of speech representation models \cite{yang2022knowledge,peng2023dphubert,huang2023ensemble,zampierin2024skill} and a much higher reduction in size than in pruning studies on speech representation models \cite{peng2023structured,pasad2023comparative,dorszeswki2024pruning}. Although our method does not bring significant improvements over other state-of-the-art distillation methods, it demonstrates the high redundancy of the transformer stack and highlights the potential for significant reductions in size and computational load.

Interestingly, we do not observe a significant difference in performance between using one or two mimicking layers, indicating that intermediate representations are not essential for the downstream task. We found that increasing the hidden dimensionality of the mimicking layer ($z$) slightly improved performance, although $z = 4096$ often led to model overfitting. While transformer mimicker layers generally produced the best results, using a single transformer or linear layer without mimicking steps also demonstrated impressive performance, suggesting the exact representations learned by the transformer stack to be non-critical. However, eliminating the transformer stack and relying solely on the final linear layer for classification resulted in a substantial performance drop to $79\%$ accuracy, implying some degree of non-linearity is required to maintain performance. These findings suggest that the transformer stack is not needed for downstream applications and can be replaced by a single non-linear layer. The redundancy of the transformer stack has also been found by Kostas et al. \cite{kostas2021bendr} in an EEG transformer model, where the transformer stack was beneficial for pretraining but not critical or even harmful for downstream applications. 

\section{Conclusion}
\label{sec:conclusion}
Our findings indicate a significant degree of redundancy within the transformer layers of speech representation models. This redundancy is evident from the high similarity between layers, particularly within the two primary blocks identified in our analysis. The ability to prune 15-45\% of transformer layers without retraining while mostly maintaining performance further underscores the significant redundancy present in transformer layers within speech representation models. The two main blocks found in the similarity analysis seem to be critical for performance, as fully pruning either block results in a massive drop in performance. However, within these blocks many layers can be pruned, suggesting high redundancy within each block.

Our exploration of mimicking networks suggests that the entire transformer stack can be replaced with a much smaller and faster network, while maintaining over 95\% of performance, highlighting the high redundancy of the transformer stack in speech representation models. This potential of reducing the model to a model order of magnitudes smaller and faster than the original model is also supported by recent studies on KD in speech representation models \cite{yang2022knowledge,peng2023dphubert,huang2023ensemble,zampierin2024skill}. All of these findings highlight the potential for substantial reduction in size and computational load enabling practical applications in on-device automatic speech recognition and resource-constrained environments.

\bibliographystyle{IEEEbib}
\bibliography{strings,refs}

\newpage
\appendix
\onecolumn
\section{Appendix / supplemental material}
\label{appendix:A}

We present the data behind \autoref{fig:mimic} (see \autoref{tab:results-wav2vec}) along with experiments of mimicking networks for the \texttt{wav2vec2-large}, \texttt{wavLM-small} and \texttt{wavLM-large}. Note that the L and T respectively denote linear and transformer layers, while $z$ is the hidden dimension of the layer(s). 

\begin{figure}[hp!]
    \centering
    \includegraphics[width=\linewidth]{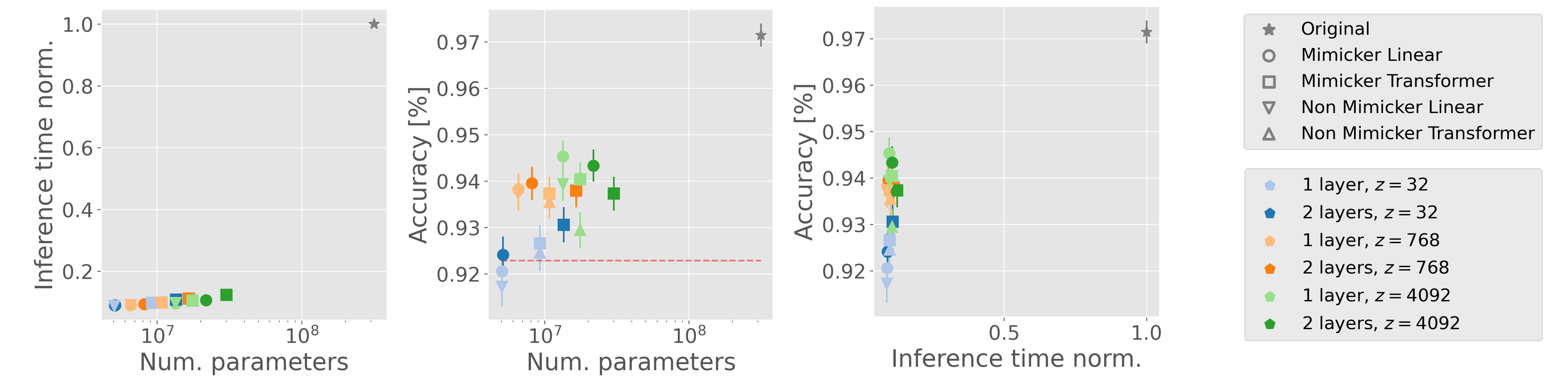}
    \caption{\texttt{wav2vec2-large}}
    \label{fig:mimic_wav2veclarge}
\end{figure}

\begin{figure}[hp!]
    \centering
    \includegraphics[width=\linewidth]{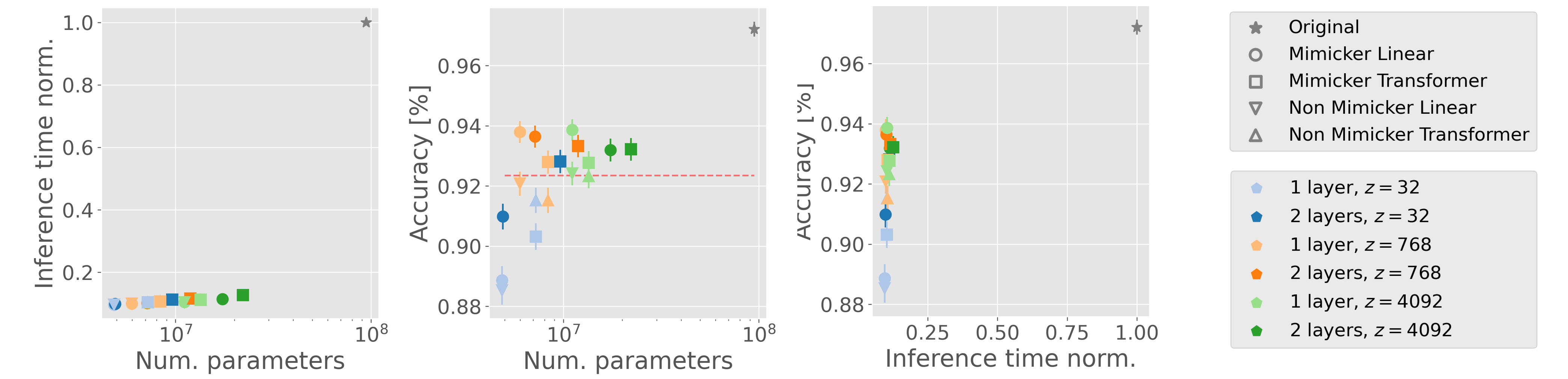}
    \caption{\texttt{wavLM-small}}
    \label{fig:mimic_wavlm}
\end{figure}

\begin{figure}[hp!]
    \centering
    \includegraphics[width=\linewidth]{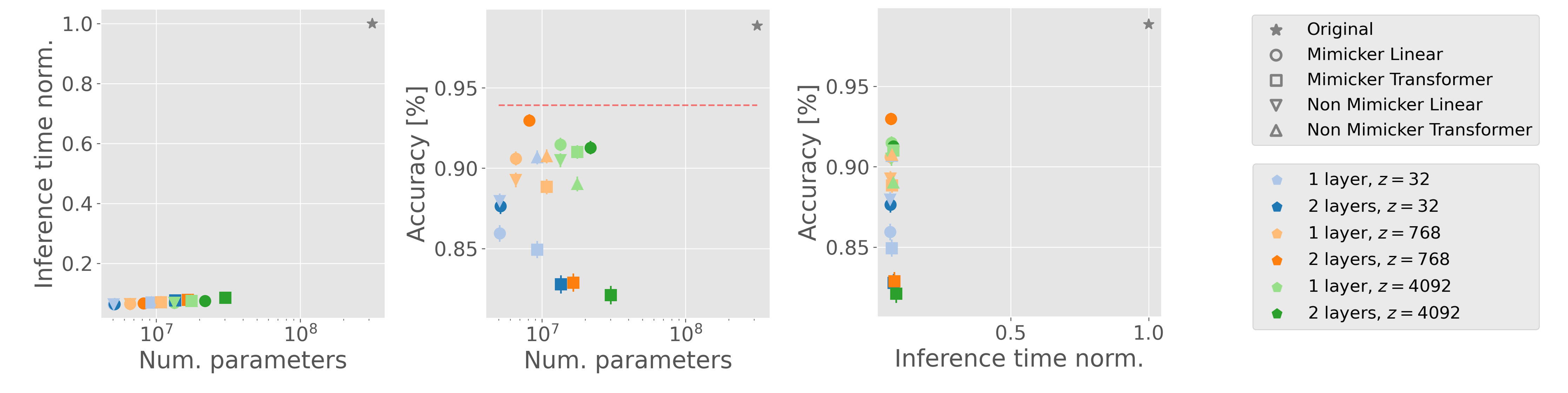}
    \caption{\texttt{wavLM-large}}
    \label{fig:mimic_wavlm-large}
\end{figure}

Results of simplification of networks using \textit{mimicking networks}. In \texttt{wav2vec2-large} (\autoref{fig:mimic_wav2veclarge} and \autoref{tab:results-wav2vec-large}) and \texttt{wavLM-small} (\autoref{fig:mimic_wavlm} and \autoref{tab:results-wavLM-small}) the models keep over 95\% of their original performance while reducing the number of parameters by 95-98\% and the inference time by up to 91\%. In \texttt{wavLM-large} (\autoref{fig:mimic_wavlm-large} and \autoref{tab:results-wavLM-large}) the performance is still above 90\% of the original performance while reducing the size by 98\% and the inference time by 94\%. 

\begin{table}[hp!]
    \centering
    \begin{tabular}{crrrrrr}
    \toprule \toprule
    Network & Layer & $N$ &  $z$ & Number of &  Inference time & Accuracy \\
    type & type & layers & & parameters & (normalized) & \\
    \midrule
        Original & T & 12 &    - & 94577571 & 1   & $0.976 \pm 0.002$ \\ \hline \rule{0pt}{2.2ex}
        Mimicker & L & 1 &   32 & 4851331 & 0.13  & $0.919 \pm 0.004$ \\
        Mimicker & L & 2 &   32 & 4901283 & 0.14  & $0.925 \pm 0.004$ \\
        Mimicker & L & 1 &  768 & 5984035 & 0.14  & $0.942 \pm 0.003$ \\
        Mimicker & L & 2 &  768 & 7165219 & 0.14  & $0.938 \pm 0.004$ \\
        Mimicker & L & 1 & 4096 & 11105827 & 0.15 & $0.935 \pm 0.004$ \\
        Mimicker & L & 2 & 4096 & 17402147 & 0.16 & $0.934 \pm 0.004$ \\ \rule{0pt}{3.5ex}
        Mimicker & T & 1 &   32 & 7216707 & 0.14  & $0.936 \pm 0.004$ \\
        Mimicker & T & 2 &   32 & 9632099 & 0.16  & $0.934 \pm 0.004$ \\
        Mimicker & T & 1 &  768 & 8347939 & 0.15  & $0.921 \pm 0.004$ \\
        Mimicker & T & 2 &  768 & 11894563 & 0.16 & $0.942 \pm 0.003$ \\
        Mimicker & T & 1 & 4096 & 13463075 & 0.16 & \textbf{$0.945 \pm 0.003$} \\
        Mimicker & T & 2 & 4096 & 22124835 & 0.18 & $0.944 \pm 0.003$ \\ \hline \rule{0pt}{2.2ex}
    Non-mimicker & L & 1 &   32 & 4851331 & 0.13  & $0.916 \pm 0.004$ \\
    Non-mimicker & L & 1 &  768 & 5984035 & 0.14  & $0.936 \pm 0.004$ \\
    Non-mimicker & L & 1 & 4096 & 11105827 & 0.14 & $\mathbf{0.945 \pm 0.003}$ \\ \rule{0pt}{3.5ex}
    Non-mimicker & T & 1 &   32 & 7216707 & 0.14  & $ 0.93 \pm 0.004$ \\
    Non-mimicker & T & 1 &  768 & 8347939 & 0.15  & $0.931 \pm 0.004$ \\
    Non-mimicker & T & 1 & 4096 & 13463075 & 0.16 & $0.925 \pm 0.004$ \\
    \bottomrule \bottomrule
    \end{tabular}
    \vspace{0.4cm}
    \caption{\texttt{wav2vec2-small}}
    \label{tab:results-wav2vec}
\end{table}

\begin{table}[hp!]
    \centering
    \begin{tabular}{crrrrrr}
    \toprule \toprule
    Network & Layer & $N$ &  $z$ & Number of &  Inference time & Accuracy \\
    type & type & layers & & parameters & (normalized) & \\
    \midrule
    Original & T & 24 &    - & 315700387 & 1   & $0.971 \pm 0.002$ \\ \hline \rule{0pt}{2.2ex}
    Mimicker & L & 1 &   32 & 5064835 & 0.089  & $0.921 \pm 0.004$ \\
    Mimicker & L & 2 &   32 & 5131427 & 0.091  & $0.924 \pm 0.004$ \\
    Mimicker & L & 1 &  768 & 6574371 & 0.09   & $0.938 \pm 0.004$ \\
    Mimicker & L & 2 &  768 & 8149027 & 0.094  & $ 0.94 \pm 0.004$ \\
    Mimicker & L & 1 & 4096 & 13400099 & 0.097 & $\mathbf{0.945 \pm 0.003}$ \\
    Mimicker & L & 2 & 4096 & 21793827 & 0.11  & $0.943 \pm 0.003$ \\ \rule{0pt}{3.5ex}
    Mimicker & T & 1 &   32 & 9267267 & 0.098  & $0.927 \pm 0.004$ \\
    Mimicker & T & 2 &   32 & 13536355 & 0.11  & $0.931 \pm 0.004$ \\
    Mimicker & T & 1 &  768 & 10775331 & 0.099 & $0.937 \pm 0.004$ \\
    Mimicker & T & 2 &  768 & 16552483 & 0.11  & $0.938 \pm 0.004$ \\
    Mimicker & T & 1 & 4096 & 17594403 & 0.11  & $ 0.94 \pm 0.004$ \\
    Mimicker & T & 2 & 4096 & 30190627 & 0.12  & $0.937 \pm 0.004$ \\ \hline \rule{0pt}{2.2ex}
Non-mimicker & L & 1 &   32 & 5064835 & 0.088  & $0.917 \pm 0.004$ \\
Non-mimicker & L & 1 &  768 & 6574371 & 0.09   & $0.937 \pm 0.004$ \\
Non-mimicker & L & 1 & 4096 & 13400099 & 0.097 & $0.939 \pm 0.004$ \\ \rule{0pt}{3.5ex}
Non-mimicker & T & 1 &   32 & 9267267 & 0.098  & $0.925 \pm 0.004$ \\
Non-mimicker & T & 1 &  768 & 10775331 & 0.1   & $0.936 \pm 0.004$ \\
Non-mimicker & T & 1 & 4096 & 17594403 & 0.11  & $0.929 \pm 0.004$ \\

    \bottomrule \bottomrule
    \end{tabular}
    \vspace{0.4cm}
    \caption{\texttt{wav2vec2-large}}
    \label{tab:results-wav2vec-large}
\end{table}

\begin{table}[hp!]
    \centering
    \begin{tabular}{crrrrrr}
    \toprule \toprule
    Network & Layer & $N$ &  $z$ & Number of &  Inference time & Accuracy \\
    type & type & layers & & parameters & (normalized) & \\
    \midrule
    Original & T & 12 &    - & 94587795 & 1   & $0.972 \pm 0.002$ \\ \hline \rule{0pt}{2.2ex}
    Mimicker & L & 1 &   32 & 4851331 & 0.097 & $0.889 \pm 0.005$ \\
    Mimicker & L & 2 &   32 & 4901283 & 0.099 & $ 0.91 \pm 0.004$ \\
    Mimicker & L & 1 &  768 & 5984035 & 0.1   & $0.938 \pm 0.004$ \\
    Mimicker & L & 2 &  768 & 7165219 & 0.1   & $0.936 \pm 0.004$ \\
    Mimicker & L & 1 & 4096 & 11105827 & 0.1  & $\mathbf{0.939 \pm 0.004}$ \\
    Mimicker & L & 2 & 4096 & 17402147 & 0.11 & $0.932 \pm 0.004$ \\ \rule{0pt}{3.5ex}
    Mimicker & T & 1 &   32 & 7216707 & 0.1   & $0.903 \pm 0.004$ \\
    Mimicker & T & 2 &   32 & 9632099 & 0.11  & $0.928 \pm 0.004$ \\
    Mimicker & T & 1 &  768 & 8347939 & 0.11  & $0.928 \pm 0.004$ \\
    Mimicker & T & 2 &  768 & 11894563 & 0.12 & $0.933 \pm 0.004$ \\
    Mimicker & T & 1 & 4096 & 13463075 & 0.11 & $0.928 \pm 0.004$ \\
    Mimicker & T & 2 & 4096 & 22124835 & 0.13 & $0.932 \pm 0.004$ \\ \hline \rule{0pt}{2.2ex}
Non-mimicker & L & 1 &   32 & 4851331 & 0.097 & $0.885 \pm 0.005$ \\
Non-mimicker & L & 1 &  768 & 5984035 & 0.1   & $0.921 \pm 0.004$ \\
Non-mimicker & L & 1 & 4096 & 11105827 & 0.1  & $0.924 \pm 0.004$ \\ \rule{0pt}{3.5ex}
Non-mimicker & T & 1 &   32 & 7216707 & 0.1   & $0.915 \pm 0.004$ \\
Non-mimicker & T & 1 &  768 & 8347939 & 0.11  & $0.915 \pm 0.004$ \\
Non-mimicker & T & 1 & 4096 & 13463075 & 0.11 & $0.923 \pm 0.004$ \\

    \bottomrule \bottomrule
    \end{tabular}
    \vspace{0.4cm}
    \caption{\texttt{wavLM-small}}
    \label{tab:results-wavLM-small}
\end{table}

\begin{table}[hp!]
    \centering
    \begin{tabular}{crrrrrr}
    \toprule \toprule
    Network & Layer & $N$ &  $z$ & Number of &  Inference time & Accuracy \\
    type & type & layers & & parameters & (normalized) & \\
    \midrule
    Original & T & 24 &    - & 315724515 & 1   & $0.989 \pm 0.002$ \\ \hline \rule{0pt}{2.2ex}
    Mimicker & L & 1 &   32 & 5070979 & 0.064  & $0.859 \pm 0.005$ \\
    Mimicker & L & 2 &   32 & 5137571 & 0.065  & $0.876 \pm 0.005$ \\
    Mimicker & L & 1 &  768 & 6580515 & 0.065  & $0.906 \pm 0.004$ \\
    Mimicker & L & 2 &  768 & 8155171 & 0.067  & $\mathbf{0.93 \pm 0.004}$ \\
    Mimicker & L & 1 & 4096 & 13406243 & 0.069 & $0.915 \pm 0.004$ \\
    Mimicker & L & 2 & 4096 & 21799971 & 0.075 & $0.913 \pm 0.004$ \\ \rule{0pt}{3.5ex}
    Mimicker & T & 1 &   32 & 9273411 & 0.07   & $0.849 \pm 0.005$ \\
    Mimicker & T & 2 &   32 & 13542499 & 0.076 & $0.828 \pm 0.006$ \\
    Mimicker & T & 1 &  768 & 10781475 & 0.07  & $0.888 \pm 0.005$ \\
    Mimicker & T & 2 &  768 & 16558627 & 0.079 & $0.829 \pm 0.006$ \\
    Mimicker & T & 1 & 4096 & 17600547 & 0.075 & $ 0.91 \pm 0.004$ \\
    Mimicker & T & 2 & 4096 & 30196771 & 0.086 & $0.821 \pm 0.006$ \\  \hline \rule{0pt}{2.2ex}
Non-mimicker & L & 1 &   32 & 5070979 & 0.064  & $ 0.88 \pm 0.005$ \\
Non-mimicker & L & 1 &  768 & 6580515 & 0.064  & $0.893 \pm 0.005$ \\
Non-mimicker & L & 1 & 4096 & 13406243 & 0.069 & $0.905 \pm 0.004$ \\  \rule{0pt}{3.5ex}
Non-mimicker & T & 1 &   32 & 9273411 & 0.07   & $0.907 \pm 0.004$ \\
Non-mimicker & T & 1 &  768 & 10781475 & 0.071 & $0.907 \pm 0.004$ \\
Non-mimicker & T & 1 & 4096 & 17600547 & 0.075 & $ 0.89 \pm 0.005$ \\
    \bottomrule \bottomrule
    \end{tabular}
    \vspace{0.4cm}
    \caption{\texttt{wavLM-large}}
    \label{tab:results-wavLM-large}
\end{table}

\end{document}